\font\titleftb=cmbx12 at 17.28pt 
\font\sectionft=cmr12 
\font\sectionftb=cmbx12 
\font\smallft=cmr9 
\font\smallbft=cmbx9 
\nopagenumbers 
\hsize=126mm 
\vsize=190mm 
\hoffset=15mm 
\voffset=22.6mm 
 
\vbox to 18pt{} 
\centerline{\titleftb Cosmic Ray Acceleration at} 
\medskip 
\centerline{\titleftb Relativistic Shock Waves} 
 
\bigskip\bigskip\bigskip 
\centerline{\sectionft  M. Ostrowski} 
\medskip 
\centerline{\it Obserwatorium Astronomiczne, Uniwersytet Jagiello\'nski,} 
\centerline{\it Krak\'ow, Poland (E-mail:  mio{@}oa.uj.edu.pl)} 
 
\bigskip\bigskip\bigskip 
\midinsert 
\centerline{\sectionftb  Abstract} \par 
\narrower 
\noindent 
{\smallft Theory of the first-order Fermi acceleration of cosmic ray 
particles at relativistic shock waves is reviewed. We consider shocks 
with  parallel and oblique, sub- and super-luminal magnetic field 
configurations and with finite-amplitude magnetic field perturbations. 
A role of oblique field configurations and field perturbations in 
forming the cosmic ray energy spectrum and changing the acceleration 
time scale is discussed.} \endinsert 
 
\bigskip 
\noindent 
{\sectionftb 1. Particle acceleration at non-relativistic shock waves} 
\medskip 
 
Processes of first-order particle acceleration at non-relativistic shock 
waves were widely discussed by a number of authors during the last two 
decades (for review, see, e.g. Drury (1983), Blandford \& Eichler 
(1987), Berezko et al. (1988), Jones \& Ellison (1991)). Below, we 
remind the basic physical picture and some most of the important results 
obtained within this theory, to be later compared with the results 
obtained for relativistic shocks. 
 
The preferred by us simple description of the acceleration process 
consists in considering two plasma rest frames, the {\it upstream frame} 
and the {\it downstream one}. We use indices `$1$' or `$2$' to indicate 
quantities measured in the upstream or the downstream frame 
respectively. If one neglects the second-order Fermi acceleration, 
the particle energy is a constant of motion in any plasma rest frame and 
energy changes occur when the particle momentum is  Lorentz-transformed 
at each crossing of the shock. In the case of {\it parallel} shock, 
with the mean magnetic field parallel to the shock normal, the 
acceleration of an individual particle is due to a consecutive 
shock crossings by the diffusively wandering particle. Each {\it 
upstream-downstream-upstream} diffusive loop results in a small 
increment of particle momentum, $\Delta p \propto U_1/v$, where $v$ is 
the particle velocity and $U_1$ is the shock velocity in the upstream 
frame, $U_1 \ll v \approx c$. In oblique shocks, the particle helical 
trajectories can cross the shock surface a number of times at any 
individual shock transition or reflection. 
 
The most interesting feature of the first-order Fermi acceleration at a 
non-relativistic plane-parallel shock wave is independence of the 
{\it test-particle stationary} particle energy spectrum from the 
background conditions near the shock, including the mean magnetic field 
configuration and the spectrum of MHD turbulence. The main reason behind 
that is a nearly-isotropic form of the particle momentum distribution at 
the shock. If a sufficient amount of scattering occurs near the shock, 
this condition always holds for the shock velocity along the upstream 
magnetic field $U_{B,1} \equiv U_1 / \cos \Psi_1 \ll v$ ($\Psi_1$ - the 
upstream magnetic field inclination to the shock normal). Independently 
of the field inclination at the shock, the particle density is 
continuous across it and the spectral index for the phase-space 
distribution function, $\alpha$, is defined exclusively in the terms of 
the shock compression ratio $R$ : 
 
$$\alpha = {3R \over R-1}  \qquad   . \eqno(1.1) $$ 
 
\noindent 
Because of the isotropic form of the particle distribution function, the 
spatial diffusion equation has become a widely used mathematical tool for 
describing particle transport and acceleration processes in 
non-relativistic flows. The characteristic acceleration time scale at 
the parallel shock is 
 
$$T_{acc} = {3 \over U_1-U_2}\, \left\{ {\kappa_1 \over 
U_1} + {\kappa_2 \over U_2} \right\} \qquad , \eqno(1.2) $$ 
 
\noindent 
where $\kappa_i \equiv \kappa_{\parallel ,i}$ is the respective particle 
spatial diffusion coefficient along the magnetic field, as discussed by 
e.g. Lagage \& Cesarsky (1983). Ostrowski (1988, see also Bednarz \& 
Ostrowski 1996) derived an analogous expression for shocks with oblique 
magnetic fields and small amplitude magnetic field perturbations. For 
a negligible cross-field diffusion and for $U_{B,1} \ll c$ it can be 
written in essentially the same form as the one given in Equ.~(1.2), 
with all quantities taken as the normal ($n$) ones with respect to the 
shock ($\kappa_{n,i}$ for $\kappa_i$ ($i$ = $1$, $2$)). Usually 
$\kappa_n < \kappa_\parallel$. Therefore, as confirmed by measurements 
in the heliosphere, the oblique shocks may be much more rapid 
accelerators as compared to the parallel shocks. 
 
\bigskip 
\noindent 
{\sectionftb 2. Cosmic ray acceleration at relativistic shock waves} 
\bigskip 
\noindent 
{\sectionftb 2.1 The Fokker-Planck description of the acceleration process} 
\medskip 
 
In the case of the shock velocity (or its projection $U_{B,1}$) reaching 
values comparable to the light velocity, the particle distribution at the 
shock becomes anisotropic. This fact complicates to a great 
extent both the physical picture and the mathematical description of 
particle acceleration. The first attempt to consider the acceleration 
process at the relativistic shock was presented in 1981 by Peacock (see 
also Webb 1985); however, no consistent theory was proposed until a 
paper of Kirk \& Schneider (1987a; see also Kirk 1988) appeared. Those 
authors considered the stationary solutions of the relativistic 
Fokker-Planck equation for particle pitch-angle diffusion for the case 
of parallel shock wave. In the situation with the gyro-phase averaged 
distribution $f(p, \mu, z)$, which depends only on the unique spatial 
co-ordinate $z$ along the shock velocity, and with $\mu$ being the 
pitch-angle cosine, the equation takes the form 
 
$$\Gamma ( U + v \mu ) {\partial f \over \partial z} = C(f) + S \qquad , 
 \eqno(2.1)  $$ 
 
\noindent 
where $\Gamma \equiv 1/\sqrt{1-U^2}$ is the flow Lorentz factor, $C(f)$ 
is the collision operator and $S$ is the source function. In the 
presented approach, the spatial co-ordinates are measured in the shock 
rest frame, while the particle momentum co-ordinates and the collision 
operator are given in the respective plasma rest frame. For the applied 
pitch-angle diffusion operator, $C = \partial / \partial \mu 
(D_{\mu,\mu} \partial f / \partial \mu)$, they generalised the diffusive 
approach to higher order terms in particle distribution anisotropy and 
constructed general solutions at both sides of the shock which involved 
solutions of the eigenvalue problem. By matching two solutions at the 
shock, the spectral index of the resulting power-law particle 
distribution can be found by taking into account a sufficiently large 
number of eigenfunctions. The same procedure yields the particle angular 
distribution and the spatial density distribution. The low-order truncation 
in this approach corresponds to the standard diffusion approximation and 
to a somewhat more general method described by Peacock. The above 
analytic approach (or the `semi-analytic' one, as the mentioned 
matching of two series involves numerical fitting of the respective 
coefficients) was verified by Kirk \& Schneider (1987b) by the method 
of particle Monte Carlo simulations. 
 
\topinsert 
\vskip 7cm  \noindent 
{\smallbft Fig. 1} 
{\smallft The particle spectral indices $\alpha$ at parallel shock waves 
propagating in the cold ($e$, $p$) plasma versus the shock velocity 
$U_1$ (Heavens \& Drury 1988). On the right vertical axis the respective 
synchrotron spectral index $\gamma$ is given. Using the solid line and 
the dashed line (long dashes) we show indices for two choices of the 
turbulence spectrum. The dashed line with short dashes gives the 
spectral index derived formally from the non-relativistic Equ.~1.1. The 
horizontal line $\alpha = 4.0$ is given for the reference purposes.} 
\endinsert 
 
An application of this approach to more realistic conditions -- but 
still for parallel shocks -- was presented by Heavens \& Drury (1988), 
who investigated the fluid dynamics of relativistic shocks (cf. also 
Ellison \& Reynolds 1991) and used the results to calculate spectral 
indices for accelerated particles (Fig.~1). They considered the parallel 
shock wave propagating into electron-proton or electron-positron plasma, 
and performed calculations using the analytic method of Kirk \& Schneider 
for two different power spectra for the scattering MHD waves. In 
contrast to the non-relativistic case, they found (see also Kirk 1988) 
that the particle spectral index depends on the form of the wave spectrum. 
An interesting fact was revealed that {\it synchrotron} spectral indices 
$\gamma$ ($\equiv (\alpha-3)/2$) obtained for shock velocities ranging 
from non-relativistic ones up to $U_1 = 0.98\,c$ fell into a relatively 
narrow gap, between $0.35$ and $0.6$ for both the fluids considered. 
They also noted a strange fact that the non-relativistic expression 
(1.1) provided a quite reasonable approximation to the actual spectral 
index. 
 
\topinsert 
\vskip 7cm  \noindent 
{\smallbft Fig. 2} 
{\smallft Spectral indices $\alpha$ of particles accelerated at oblique 
shocks versus shock velocity projected at the mean magnetic field, 
$U_{B,1}$. On the right vertical axis the respective synchrotron 
spectral index $\gamma$ is given. The shock velocities $U_1$ are given 
near the respective curves taken from Kirk \& Heavens (1989). The points 
were taken from simulations deriving explicitly the details of 
particle-shock interactions (Ostrowski 1991a). The results are presented 
for compression $R = 4$.} \endinsert 
 
A substantial progress in understanding the acceleration process in the 
presence of highly anisotropic particle distributions is due to the work 
of Kirk \& Heavens (1989; see also a discussion in Ostrowski 1991a and 
Ballard \& Heavens 1991), who considered particle acceleration at {\it 
subluminal} ($U_{B,1} < c$) relativistic shocks with oblique magnetic 
fields. They assumed the magnetic momentum conservation, $p_\perp^2/B = 
const$, at particle interaction with the shock and applied the 
Fokker-Planck equation discussed above to describe particle transport 
along the field lines outside the shock. Within the considered approach, 
a possibility of cross-field diffusion is excluded. In the cases when 
$U_{B,1}$ reached relativistic values, they derived very flat energy 
spectra with $\gamma \approx 0$ at $U_{B,1} \approx 1$ (Fig.~2). In such 
conditions particle density in front of the shock can substantially - 
even by a few orders of magnitude - exceed the downstream density (see 
the curve denoted `-8.9' at Fig.~3). Creating flat spectra and great 
density contrasts is due to effective reflections of anisotropically 
distributed upstream particles from the region of compressed magnetic 
field downstream the shock. However, the conditions leading to very flat 
spectra are supposed to be accompanied by processes - like a large 
amplitude wave generation upstream the shock - leading to the spectrum 
steepening (Ostrowski 1991a; see also Sec.~2.3). 
 
As stressed by Begelman \& Kirk (1990), in relativistic shocks one can 
often find the superluminal conditions, with $U_{B,1} > c$, where the 
above presented approach is no longer valid. Then, it is not possible to 
reflect upstream particles from the shock and to transmit downstream 
particles into the upstream region. In effect, only a single compression 
of the transmitted particles' helical orbits re-shapes the upstream 
particle distribution by shifting particle energies to larger values. 
Begelman \& Kirk show that the energy gains in such a process can be 
quite significant, exceeding the value expected for the adiabatic 
compression. The physical reason behind that is the growing particle 
anisotropy during compression at the shock, providing the excessive 
pressure along the shock normal. 
 
The approach proposed by Kirk \& Schneider (1987a) and the derivations 
of Begelman \& Kirk (1990) are valid only in the case of weakly 
disturbed background magnetic fields. However, in the efficiently 
accelerating shocks one may expect the large amplitude waves to be 
present, when both the Fokker-Planck approach is no longer valid and the 
magnetic momentum conservation no longer holds for oblique shocks 
(Ostrowski 1991a). In such a case, numerical methods have to be used. 
 
\bigskip 
\noindent 
{\sectionftb 2.2 Particle acceleration in the presence of large amplitude 
magnetic field perturbations} 
\medskip 
 
\topinsert 
\vskip 7cm  \noindent 
{\smallbft Fig. 3} 
{\smallft The energetic particle density across the relativistic shock with 
oblique magnetic field. The shock with $U_1 = 0.5$, $R = 5.11$ and 
$\psi_1 = 55^o$ is considered. The curves for different perturbation 
amplitudes are characterized with the value $\log \kappa_\perp / 
\kappa_\parallel$ given near the curve. The data are vertically shifted 
for picture clarity. The value $X_{max}$ is the distance from the shock 
at which the upstream particle density decreases to $10^{-3}$ part of 
the shock value.} \endinsert 
 
The first attempt to consider the acceleration process at parallel shock 
wave propagating in a turbulent medium was presented by Kirk \& 
Schneider (1988), who included into Equ.~2.1 the Boltzmann collision 
operator describing the large angle scattering. By solving the resulting 
integro-differential equation with the use of analytic means they 
demonstrated hardening of the particle spectrum with increasing 
contribution of the large-angle scattering. The reason for such a 
spectral change is the additional isotropization of particles 
interacting with the shock, leading to the increased particle mean 
energy gain. In oblique shocks, this simplified approach can not be 
used, not only because of the distribution anisotropy, but also due to 
the fact that the character of individual particle-shock interaction - 
reflection and transmission characteristics - depends on magnetic field 
perturbations. Let us additionally note that application of the point-like 
large angle scattering model in relativistic shocks does not provide a 
viable physical representation of the scattering at MHD waves (Bednarz 
\& Ostrowski 1996). 
 
\topinsert 
\vskip 7cm  \noindent 
{\smallbft Fig. 4} 
{\smallft Spectral indices for oblique relativistic shocks versus 
perturbation amplitude $\delta B/B$ (Ostrowski 1993). Different field 
inclinations are characterized by values of $U_{B,1}$ given near the 
respective results, $U_{B,1} < 1$ for subluminal shocks and $U_{B,1} \ge 
1$ for superluminal ones. Absence of data for small field 
amplitudes in superluminal shocks is due to no-power-law character 
of the spectrum or extremely steep power law spectra occurring in these 
conditions.} 
\endinsert 
 
To handle the problem of particle spectrum in a wide range of background 
conditions the particle Monte Carlo simulations were proposed (Kirk \& 
Schneider 1987b; Ellison et al. 1990; Ostrowski 1991a, 1993; Ballard \& 
Heavens 1992). At first, let us consider subluminal shocks. The field 
perturbations influence the acceleration process in various ways. As 
they enable the particle cross field diffusion, a modification 
(decrease) of the downstream particle's escape probability may occur. 
This factor tends to harden the spectrum. Next, the perturbations decrease 
particle anisotropy, leading to increase of the mean energy gain of 
reflected upstream particles, but - which is more important for oblique 
shocks - this increases also the particle downstream transmission 
probability, enabling them to escape efficiently from the further 
acceleration. The third factor is due to perturbing particle trajectory 
during an individual interaction with the shock discontinuity and 
breakdown of the approximate conservation of $p_\perp^2/B$. Because 
reflecting a particle from the shock requires a fine tuning 
of particle trajectory with respect to the shock surface, even a very 
small amount of perturbations can decrease the reflection probability 
in a substantial way. The influence at the acceleration time scale is 
discussed in the next section. Simulations show (Fig.~4) that - until 
the wave amplitude becomes very large - factors leading to more 
efficient particle escape are dominant and may result in steepening of 
the spectrum to $\gamma \sim 0.5 - 0.8$ may occur. At extremely large 
wave amplitudes, a slight flattening of the spectrum may take place. As 
presented at Fig.~3, the increased transmission probability for upstream 
particles leads also to lowering the cosmic ray density contrast across 
the shock (Ostrowski 1991b). 
 
In parallel shock waves propagating in a highly turbulent medium the 
effects discovered for oblique shocks can also manifest because of 
the {\it local} perturbed magnetic field compression at the shock. The 
problem was considered using the technique of particle simulations by 
Ballard \& Heavens (1992). They showed a possibility to have very steep 
spectra in this case, with the spectral index growing from $\gamma \sim 
0.6$ at medium relativistic velocities up to nearly $2.0$ at $U_1 = 
0.98$. These results apparently do not correspond to 
large-perturbation-amplitude limit of Ostrowski's (1993) simulations for 
oblique shocks and the analytic results of Heavens \& Drury (1988). In 
the paper of Ostrowski possible reasons for the discrepancy are 
discussed. 
 
For large amplitude magnetic field perturbations the acceleration 
process in the super-luminal shocks can lead to the power-law particle 
spectrum formation, against the statements of Begelman \& Kirk (1990) 
valid at small wave amplitudes only. Such general case was discussed by 
Ostrowski (1993) (Fig.~4) and by Bednarz \& Ostrowski (1996, see below) 
using the method of particle simulations. 
 
\bigskip 
\noindent 
{\sectionftb 2.3 The acceleration time scale} 
\medskip 
 
\topinsert 
\vskip 7cm  \noindent 
{\smallbft Fig. 5} 
{\smallft The acceleration time $T_{acc}$ versus the level of particle 
scattering measured by the ratio of $\kappa_\perp / \kappa_\parallel$ 
for the shock with velocity $U_1 = 0.5$. We present results for three 
values of the magnetic field inclination: a.) parallel shock ($\psi_1 = 
1^\circ$), b.) a sub-luminal shock with $\psi_1 = 45.6^\circ $ and c.) a 
super-luminal shock with $\psi_1 = 89^\circ $. The {\it 
maximum} value of the model parameter $\Delta t$ -- corresponding to the 
small wave amplitude limit -- is given at the end of each curve and the 
wave amplitude monotonously increases along each curve up to $\delta B 
\sim B$. $r_{e,1}$ is the particle gyroradius in the effective 
(including perturbations) upstream magnetic field (cf. Bednarz \& 
Ostrowski 1996).} 
\endinsert 
 
The shock waves propagating with relativistic velocities raise also 
interesting questions pertaining to the cosmic ray acceleration time 
scale, $T_{acc}$. A simple comparison to the non-relativistic formula 
based on numerical simulations shows that $T_{acc}$ relatively decreases 
with increasing shock velocity for parallel (Quenby \& Lieu 1989; 
Ellison et al. 1990) and oblique (Takahara \& Terasawa 1990; Newman et 
al. 1992; Lieu et al. 1994; Quenby \& Drolias 
1995; Naito \& Takahara 1995) shocks. However, the numerical approaches 
used there, based on assuming particle isotropization for all 
scatterings, neglect or underestimate a rather significant factor 
affecting the acceleration process -- the particle anisotropy. Ellison 
et al. (1990) and Naito \& Takahara (1995) included also the more 
realistic, in our opinion, derivations involving the pitch-angle diffusion 
approach. The calculations of Ellison et al. for parallel shocks show 
similar results to those they obtained for large amplitude 
scattering. In the shock with velocity $0.98\,c$ the acceleration time 
scale is reduced by the factor $\sim 3$ with respect to the 
non-relativistic formula of Equ.~1.2~. Naito \& Takahara considered 
shocks with oblique magnetic fields. They confirmed reduction of the 
acceleration time scale with increasing inclination of the magnetic 
field, derived earlier for non-relativistic shocks. However, their 
approach neglected effects of particle cross field diffusion and assumed 
the adiabatic invariant conservation in particle interactions with the 
shock. These two simplifications limit validity of their results to the 
cases with small amplitude turbulence near the shock. 
 
\topinsert 
\vskip 7cm  \noindent 
{\smallbft Fig. 6} 
{\smallft The values of $T_{acc}$ in units of $r_{e,1} / c$ at different 
inclinations $\psi_1$ versus the particle spectral index $\alpha$. The 
values resulting from simulations are presented for $U_1$ = $0.5$ for five 
values of the angle $\psi_1$ given near the respective results. The {\it 
maximum} value of the model parameter $\Delta t$ -- corresponding to the 
small wave amplitude limit -- is given at the end of each curve and the 
wave amplitude monotonously increases along each curve up to $\delta B 
\sim B$; $r_{e,1}$ - see Fig.~5.} \endinsert 
 
A much wider discussion of the acceleration time is provided recently by 
Bednarz \& Ostrowski (1996), who apply numerical simulations involving 
the small angle particle momentum scattering (Ostrowski 1991a). However, 
the approach is believed to provide also a reasonable description of 
particle transport in the presence of large amplitude magnetic field 
perturbations, and thus to enable modeling the effects of cross-field 
diffusion. The authors suggest that due to noticeable correlations 
present in the acceleration process, any derivation of the acceleration 
time scale cannot use the distribution of energy gains and the 
distribution of times between successive particle-shock interactions 
separately. $T_{acc}$ is defined as the time scale describing the rate 
of change of the cut-off energy in the time dependent particle spectrum 
evolution. The results are presented for shock waves with parallel and 
oblique (both, sub- and super-luminal) magnetic field configurations, 
and field perturbations with amplitudes ranging from very small ones up 
to $\delta B \sim B$ (Fig-s~5,6). In all cases the values of $T_{acc}$ 
are given in the shock {\it normal} rest frame. In parallel shocks 
$T_{acc}$ diminishes with the growing perturbation amplitude (Fig.~5) 
and the shock velocity $U_1$. However, it is approximately constant for a 
given value of $U_1$ if we use the formal diffusive time scale, 
$\kappa_1/(U_1c) + \kappa_2/(U_2c)$, as the time unit. Another 
qualitative feature discovered in oblique shocks is that due to the 
cross-field diffusion $T_{acc}$ can change with $\delta B$ in a 
non-monotonic way (Fig.~5). The acceleration process leading to the 
power-law spectrum is possible in super-luminal shocks only in the 
presence of large amplitude turbulence. Then, in contrast to the 
quasi-parallel shocks, $T_{acc}$ increases with increasing $\delta B$. 
In some cases with the oblique magnetic field configurations one may 
note a possibility to have an extremely short acceleration time scales 
comparable to the particle gyroperiod in the magnetic field upstream the 
shock. A coupling between the acceleration time scale and the particle 
spectral index is presented in Fig.~6. One should note that the form of 
involved relation is contingent to a great extent on the magnetic field 
configuration. 
 
\bigskip 
\noindent 
{\sectionftb 3. Final remarks} 
\medskip 
 
Substantial amount of work done to date on the {\it test particle} 
cosmic ray acceleration at relativistic shocks yielded not too promising 
results for meaningful modeling of the observed astrophysical radiation 
sources. The main reason for that deficiency is - in contrast to 
non-relativistic shocks - a direct dependence of the derived spectra on 
the conditions at the shock. Not only the shock compression ratio, but 
also other parameters, like the mean inclination of the magnetic field 
or the wave spectrum shape and amplitude, are significant here. 
Depending on the actual conditions one may obtain spectral indices as 
flat as $\alpha = 3.0$ ($\gamma = 0.0$) or very steep ones, $\alpha > 
5.0$ ($\gamma > 1.0$). The background conditions leading to the very 
flat spectra are probably subject to some instabilities; however, there 
is no detailed derivation describing the instability growth and the 
resulting cosmic ray spectrum modification. 
 
It seems that a true progress in modeling particle acceleration in 
actual sources requires a full non-linear description, including 
feedback of accelerated particles at the turbulent wave fields near the 
shock wave, flow modification caused by the cosmic rays' plasma 
pre-shock compression and, of course, the appropriate boundary 
conditions. A simple approach to the parallel shock case was presented 
by Baring \& Kirk (1991), who found that relativistic shocks could be 
very efficient accelerators. However, it seems to us that in a more 
general case it will be very difficult to make any substantial progress 
in that matter. The difficulty arises here from the fact that the 
generated energetic particle spectrum (and the corresponding particle 
pressure) depends on the number of parameters, and any choice of these 
parameters can substantially affect the resulting plasma flow pattern 
across the shock and the particle spectrum. Moreover, for very flat 
spectra obtained at the shock the non-linear acceleration picture 
depends to a large extent on the detailed knowledge of the background 
and boundary conditions in the scales relevant for particles near the 
upper energy cut-off. The existence of stationary solutions is doubtful 
in this case. 
 
One may note that observations of possible sites of relativistic shock 
waves (knots and hot spots in extragalactic radio sources), which allow 
for determination of the energetic electron spectra, often yield 
particle spectral indices close to $\alpha = 4.0$ ($\gamma = 0.5$). In 
order to overcome difficulties in accounting for these data Ostrowski 
(1994a) proposed an additional {\it `law of nature'} for non-linear 
cosmic ray accelerators. The particles within different energy ranges do 
not couple directly with each other and are supposed to form independent 
`degrees of freedom' in the system. Our `law' provides that the nature 
prefers energy equipartition between such degrees of freedom, yielding 
the spectra with $\alpha \approx 4.0$ . 
 
Finally, let us mention an independent possibility of particle 
acceleration in relativistic flows at {\it non-compressive} tangential 
discontinuities or shear layers (see Ostrowski in the present volume). 
This interesting possibility has not been adequately discussed to date. 
 
The present work was supported by the {\it Komitet Bada\'n Naukowych} 
(Project 2~P03D~016~11). 
 
\medskip 
\noindent 
{\sectionftb References} \par \vskip 2mm 
 
\noindent 
Ballard, K.R., Heavens, A.F., 1991. MNRAS, {\bf 251}, 438.\hfill \break 
Ballard, K.R., Heavens, A.F., 1992. MNRAS, {\bf 259}, 89. \hfill \break 
Baring, M.G., Kirk, J.G., 1990. A\&A, {\bf 241}, 329.     \hfill \break 
Bednarz, J., Ostrowski, M., 1996. MNRAS (in press). \hfill \break 
Begelman, M.C., Kirk J.G., 1990. ApJ, {\bf 353}, 66.      \hfill \break 
Berezko,E.G., Iolshin,W.K., Krymskij,G.F., Pietukhov,S.I., 
 1988, {\it Cosmic Ray \hfill \break \indent Generation at Shock Waves} 
(in Russian), Nauka, Moscow. \hfill \break 
Blandford, R.D., Eichler, D., 1987. Physics Reports, {\bf 154}, 1.\hfill \break 
Drury, L.O'C., 1983. Rep. Prog. Phys., {\bf 46}, 973.      \hfill \break 
Ellison, D.C., Jones, F.C., Reynolds, S.P., 1990. ApJ, {\bf 360}, 
702.\hfill \break 
Ellison, D.C., Reynolds, S.P., 1991. ApJ, {\bf 382}, 242.  \hfill \break 
Heavens, A., Drury, L'O.C., 1988. MNRAS, {\bf 235}, 997.   \hfill \break 
Jones, F.C., Ellison, D.C., 1991. Space Sci. Rev., {\bf 58}, 259.\hfill \break 
Kirk J.G., 1988, {\it Habilitation Theses}, preprint No.  345, 
  Max-Planck-Institut f\"ur \hfill \break \indent 
Astrophysik, Garching.         \hfill \break 
Kirk, J.G., Heavens, A., 1989. MNRAS, {\bf 239}, 995.      \hfill \break 
Kirk, J.G., Schneider, P., 1987a. ApJ, {\bf 315}, 425.     \hfill \break 
Kirk, J.G., Schneider, P., 1987b. ApJ, {\bf 322}, 256.     \hfill \break 
Kirk J.G., Schneider P., 1988, A\&A, {\bf 201}, 177.             \hfill \break 
Lagage P.O., Cesarsky C., 1983, A{\&}A, {\bf 125}, 249. \hfill \break 
Lieu, R., Quenby, J.J., Drolias, B., Naidu, K., 1994, ApJ, 
{\bf 421}, 211. \hfill \break 
Naito T., Takahara F., 1995, MNRAS, {\bf 275}, 1077. \hfill \break 
Newman P.L., Moussas X., Quenby J.J., Valdes-Galicia J.F., 
  Theodossiou-Ekateri- \hfill \break \indent 
nidi Z., 1992, A\&A, {\bf 255}, 443. \hfill \break 
Ostrowski, M., 1991a. MNRAS, {\bf 249}, 551.     \hfill \break 
Ostrowski, M., 1991b. in `{\it Relativistic Hadrons in Cosmic 
Compact Objects}', \hfill \break 
\indent eds. A. Zdziarski and M. Sikora.\hfill \break 
Ostrowski, M., 1993. MNRAS, {\bf 264}, 248.               \hfill \break 
Ostrowski, M., 1994a. Comments on Astrophysics, {\bf 17}, 207. \hfill \break 
Ostrowski M., 1994b, in Proc. 26th Meeting of the  Polish 
Astr. Soc., \hfill \break \indent 
eds. M. Sarna \& J. Zalewski, Warszawa. \hfill \break 
Peacock, J.A., 1981. MNRAS, {\bf 196}, 135.       \hfill \break 
Quenby J.J., Lieu R., 1989, Nature {\bf 342}, 654.      \hfill \break 
Quenby J.J., Drolias B., 1995, in Proc. 24th Int. Cosmic Ray Conf., 
3, 261, Rome. \hfill \break 
Takahara F., Terasawa T., 1990, in Proc. ICRR Int. Symp. 
{\it `Astrophysical Aspects \hfill \break \indent 
of the Most Energetic Cos\-mic Rays'}, eds. M. Nagano \& F. Takahara, 
Kofu. \hfill \break 
Webb, G.M., 1985. ApJ, {\bf 296}, 319. \par 
 
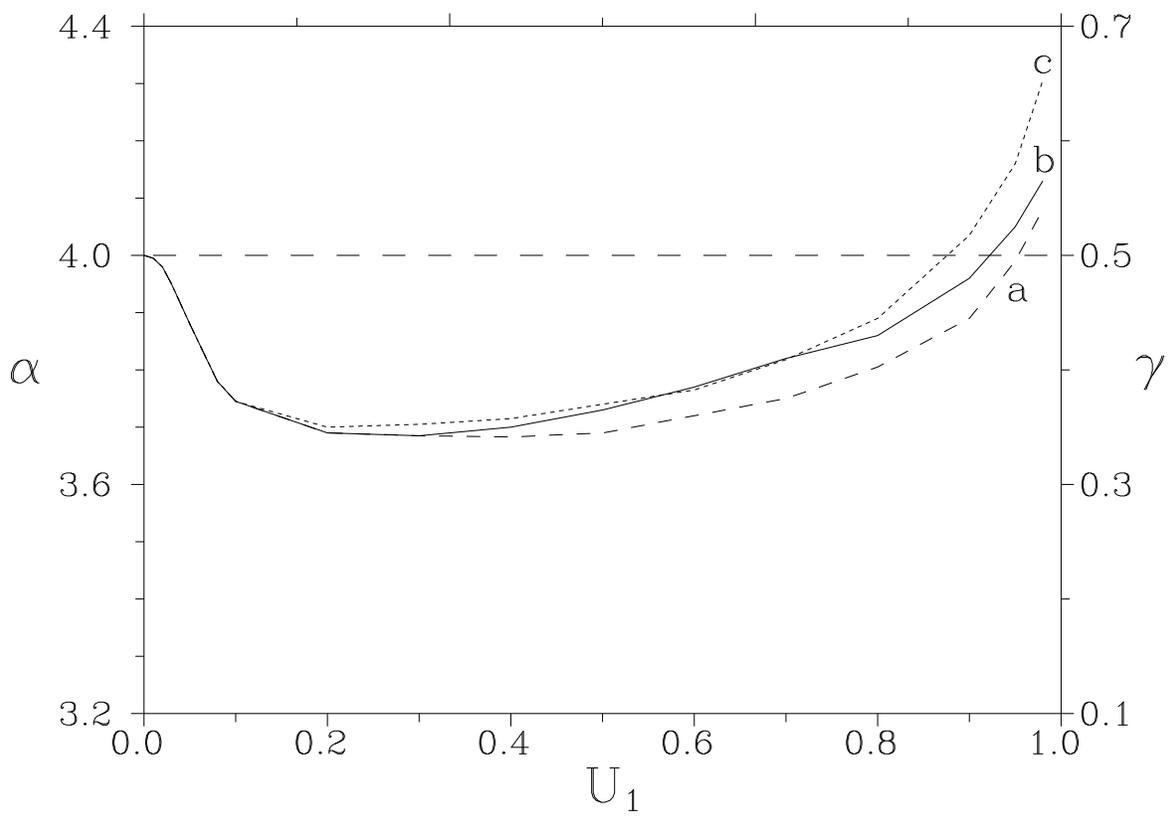
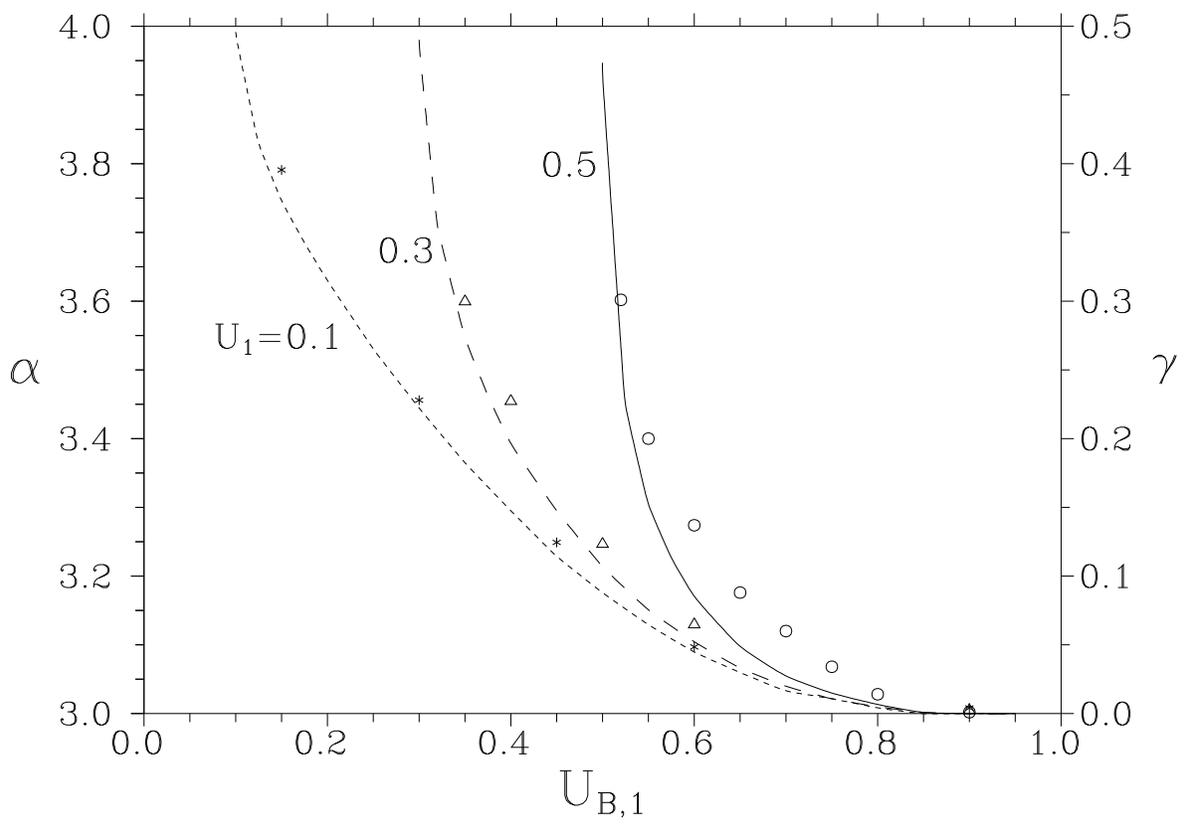
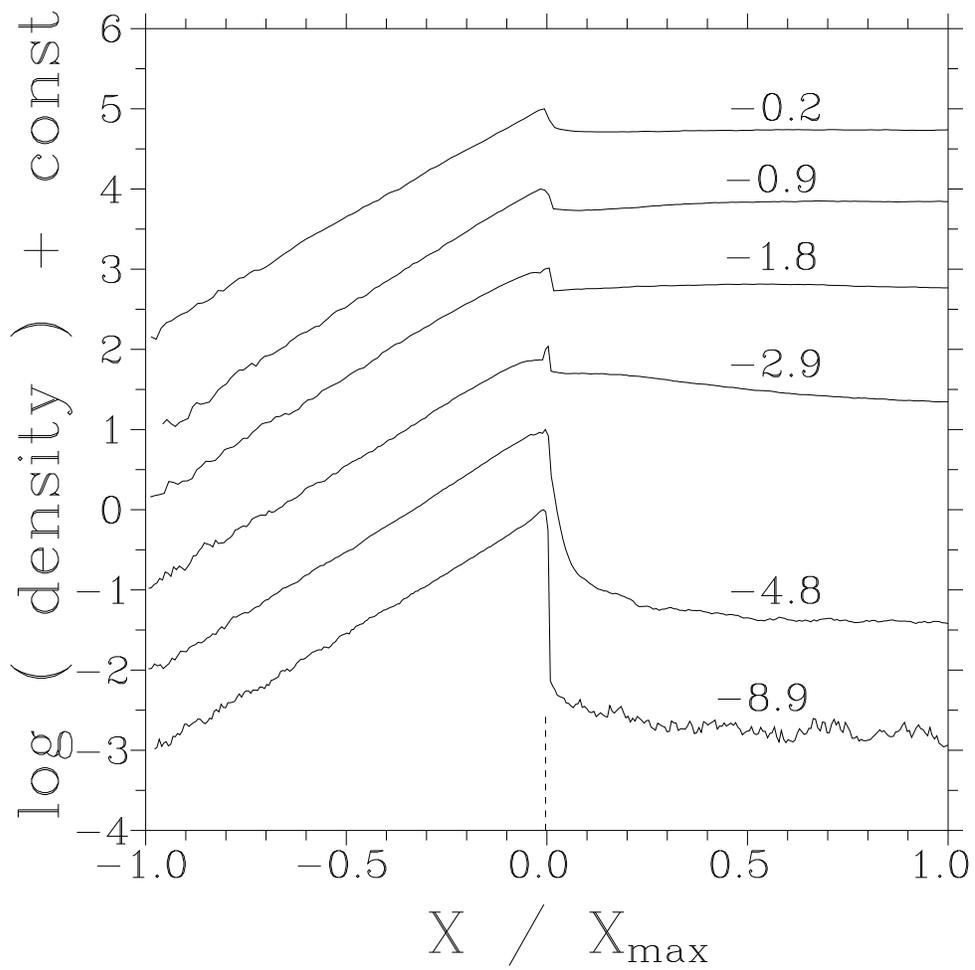
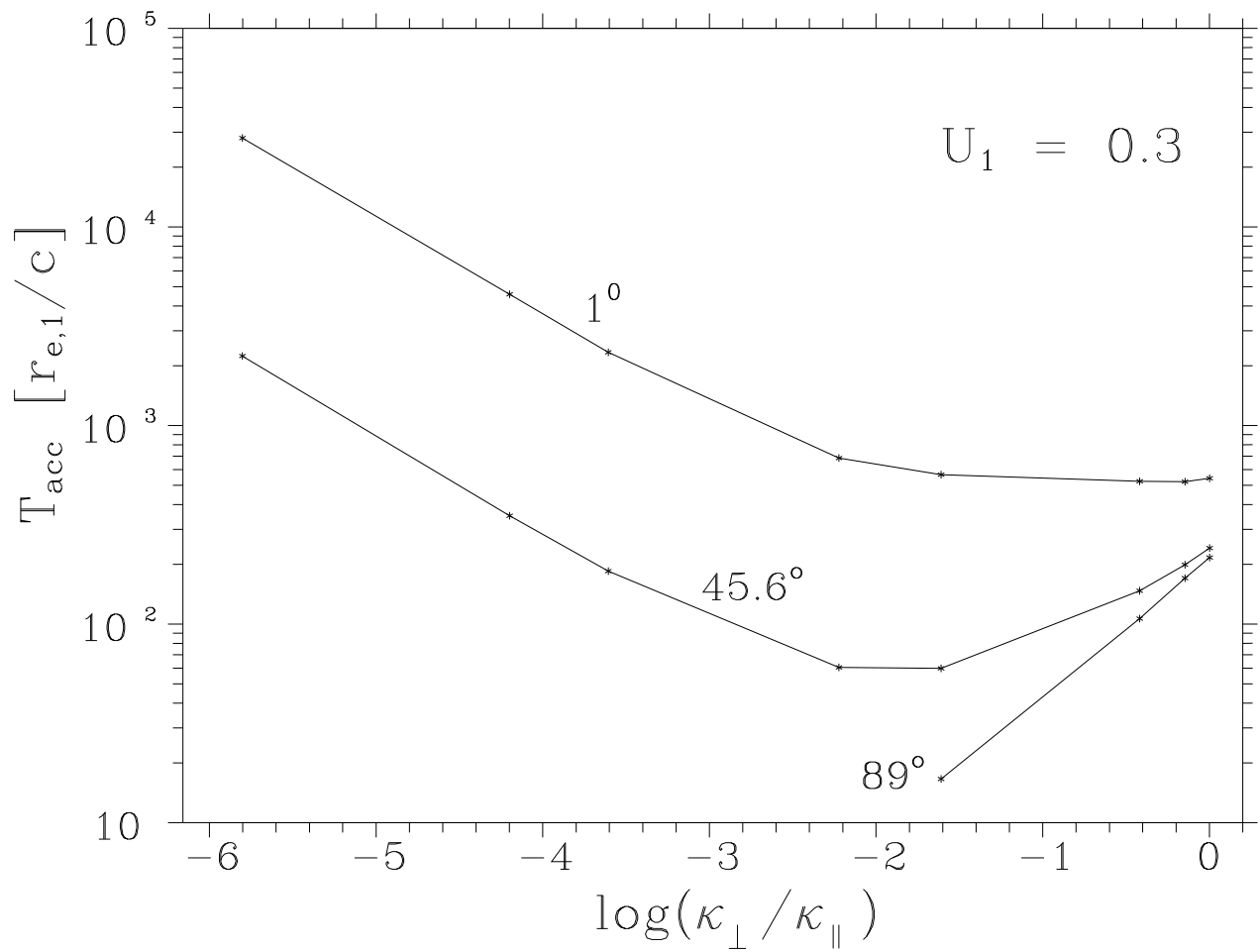
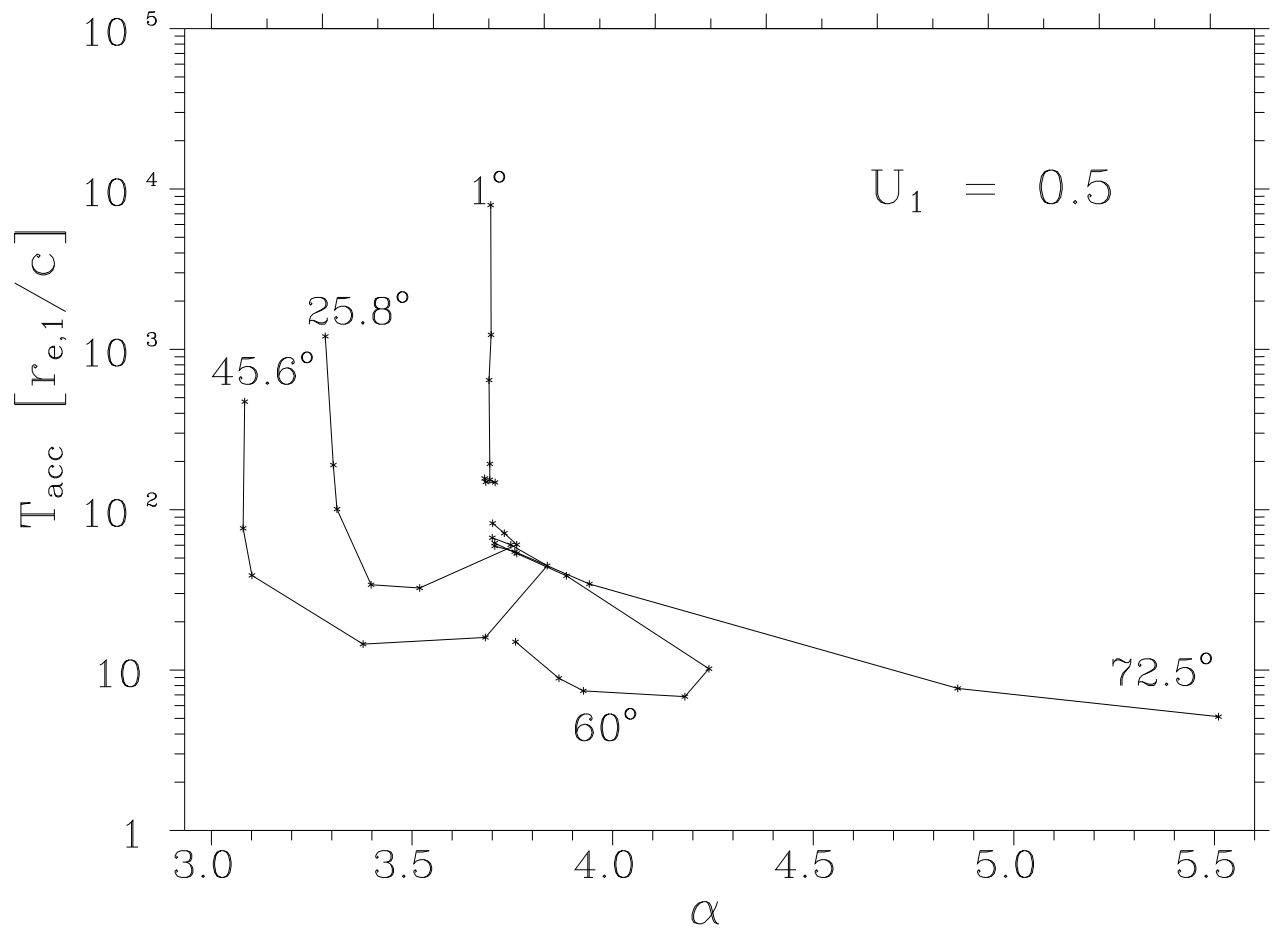
\bye